# Unique Astrophysics in the Lyman Ultraviolet

Jason Tumlinson[1], Alessandra Aloisi[1], Gerard Kriss[1], Kevin France[2], Stephan McCandliss[3], Ken Sembach[1], Andrew Fox[1], Todd Tripp[4], Edward Jenkins[5], Matthew Beasley[2], Charles Danforth[2], Michael Shull[2], John Stocke[2], Nicolas Lehner[6], Christopher Howk[6], Cynthia Froning[2], James Green[2], Cristina Oliveira[1], Alex Fullerton[1], Bill Blair[3], Jeff Kruk[7], George Sonneborn[7], Steven Penton[1], Bart Wakker[8], Xavier Prochaska[9], John Vallerga[10], Paul Scowen[11]

*Summary: There is unique and groundbreaking science to be done with a new generation of UV spectrographs that cover wavelengths in the "Lyman Ultraviolet" (LUV; 912 - 1216 Å). There is no **astrophysical** basis for truncating spectroscopic wavelength coverage anywhere between the atmospheric cutoff (3100 Å) and the Lyman limit (912 Å); the usual reasons this happens are all technical. The unique science available in the LUV includes critical problems in astrophysics ranging from the habitability of exoplanets to the reionization of the IGM. Crucially, the local Universe (z ≲ 0.1) is entirely closed to many key physical diagnostics without access to the LUV. These compelling scientific problems require overcoming these technical barriers so that future UV spectrographs can extend coverage to the Lyman limit at 912 Å.*

**The bifurcated history of the Space UV:** Much the course of space astrophysics can be traced to the optical properties of ozone ($O_3$) and magnesium fluoride ($MgF_2$). The first causes the space UV, and the second divides the "HST UV" (1150 - 3100 Å) and the "FUSE UV" (900 - 1200 Å). This short paper argues that some critical problems in astrophysics are best solved by a future generation of high-resolution UV spectrographs that observe all wavelengths between the Lyman limit and the atmospheric cutoff.

Of the major facilities of space UV astronomy (Table 1), only Copernicus and FUSE have covered the LUV, and only FUSE carried modern photon-counting detectors. IUE and HST's spectrographs (until recently) covered only λ > 1150 Å because the common $MgF_2$ optical coatings have a ~50x drop in reflectivity between 1150 and 1100 Å. COS now has limited LUV capability ($A_{eff}$ ~ 10 cm$^2$) because its design places these wavelengths onto its detector and the HST mirror coating is relatively clean after 20+ years (McCandliss et al. 2010, Osterman et al. 2010). However, this FUSE-like performance works well only for UV-bright objects and will not solve the critical problems that we describe here.

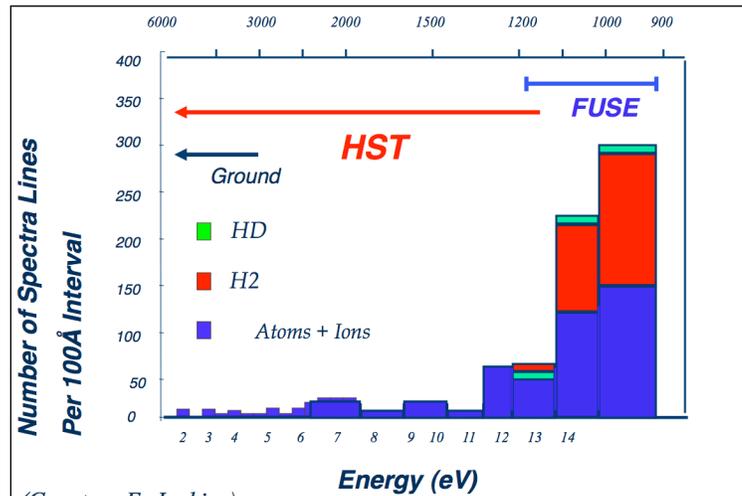

(Courtesy E. Jenkins)

Figure 1: The density of absorption-line diagnostics as a function of wavelength. The ranges for HST, FUSE and ground apply to lines observed at the **rest-frame** wavelengths at the top. Many key science cases rely on redshifting < 900 Å EUV lines into the LUV (Tripp et al. RFI response).

**The Key Reason to Cover the LUV** is the extremely rich set of unique physical diagnostics that are available there. Figure 1 shows the continuously increasing density of all spectroscopic line diagnostics toward

---

[1] STScI [2]Colorado [3]JHU [4]UMass [5]Princeton [6]Notre Dame [7]GSFC [8]Wisconsin [9]Santa Cruz [10]Berkeley [11]ASU

Table 1: UV Spectrographs and Their Wavelength Coverage

| Mission | | Lifetime | UV Coverage (Å) |
|---|---|---|---|
| Copernicus | | 1972 - 1981 | 900 - 3150 |
| IUE | | 1978 - 1996 | 1150 - 3200 |
| FUSE | | 1999 - 2007 | 900 - 1190 |
| HST | FOS | 1990 - 1997 | 1150 - 8000 |
| | GHRS | 1990 - 1997 | 1150 - 3200 |
| | STIS | 1997- now | 1150 - 3200 |
| | COS | 2009 - now | Cyc 17-19: 1140 - 3200<br>Cyc 20+: 900 - 3200 |

the LUV. The most commonly used LUV tracers include: (1) the doublet of O VI ($O^{+5}$), which traces coronal gas in the ISM and IGM, (2) the Lyman-Werner bands of molecular hydrogen, which can be detected in column densities orders of magnitude lower than those accessible to radio measurements, (3) species like Ne VIII, S VI, Fe XVIII, and Mg X, which trace gas at $\geq 10^{5-6}$ K in stellar flares, AGN outflows, and IGM gas, (4) the Lyman series itself (912 - 1216 Å), the most sensitive probe of neutral gas anywhere on the EM spectrum. These diagnostic lines provide access to gas from below 100 K to above 10 million K, and from the underdense IGM ($10^{12}$ cm$^{-2}$) to the edges of giant molecular clouds ($10^{21}$ cm$^{-2}$). *They are versatile, essential tools of astrophysics which we must observe in the LUV if we are to see them in the local Universe* (e.g. in the rest frame). We need to access these important tracers everywhere, thus our major conclusion:

**There is no *astrophysical* reason to break spectroscopic wavelength coverage anywhere between the atmospheric cutoff and the Lyman limit.**

History teaches that such breaks are driven by the availability of optical coatings and detectors, and that astrophysical goals are often compromised to conform to technical limitations of these components. We will now consider some illustrative scientific opportunities remaining untapped in the LUV, covering a wide wide range of astrophysics from exoplanets to the reionization of the IGM. A full range of additional UV science was covered in the recent "UV Astronomy: HST and Beyond" conference which featured many excellent talks now posted online (http://uvastro2012.colorado.edu/).

### UV Radiation and the Formation of Biosignatures

Stars set the conditions for life on their planets. Stellar characterization is critical to our understanding of exoplanetary life, particularly in light of the growing number of Earth-mass planets detected around low-mass stars (Batalha et al. 2012). A major source of uncertainty in the potential habitability of these worlds is the strength and variability of the local radiation fields where liquid water can persist (the "habitable zone", or HZ). Stellar flares can irradiate the exoplanetary atmosphere with heavy doses of X-ray and UV photons, potentially catalyzing or retarding the development of biology on these worlds through effects on molecular chemistry. The amplitude and frequency of UV flares on low-mass exoplanet hosts is completely unknown at present; time-resolved LUV observations are an essential input for models of habitable planets.

The large (LUV+FUV)/NUV stellar flux ratios in the HZ around M-dwarfs can have a profound effect on the atmospheric oxygen chemistry of Earth-like planets (France et al. 2012). Lyα itself is a critical energy input to the atmospheres, particularly the photochemistry of water and $CO_2$, because it contributes as much as all the rest of the

900-3000 Å range combined for stars with $T_{eff}$ ≤ 4000 K. The abiotic $O_2$ production rate (through $CO_2$ dissociation) and the subsequent formation of $O_3$ are highly dependent on the spectral and temporal behavior of the LUV, FUV, and NUV radiation field of the host stars. LUV lines from species such as O VI and FeXVIII are excellent proxies for the EUV and soft X-ray emission from low-mass stars (Redfield et al. 2003), which can estimate the energy deposition onto the planetary atmosphere without costly or impractical EUV and X-ray observations. But only three of the over 100 M-dwarfs known to host planets have well-characterized UV spectra, which hampers our ability to accurately predict the biosignatures from these worlds (Kaltenegger et al. 2011). Without direct measurements of stellar UV emission, we will not be able to assess the potential of false positives for biomarkers that may be detected in the coming decade.

With a future high-sensitivity UV mission employing photon-counting detectors and covering the LUV through the atmospheric cutoff, we will be able to survey all of the known M-dwarf exoplanet host stars at <50 pc (and K-dwarfs to > 200 pc), including all of the systems that can be studied in detail by JWST. These UV tracers provide a complete picture of the energetic radiation environment in which potentially habitable planets are immersed; directly related to the conditions for and detectability of life on these worlds.

## The Circumgalactic Medium in High Fidelity

How galaxies acquire their gas, process it, and return it to their environment as feedback are some of the most important issues in astrophysics. But the Circumgalactic Medium (CGM) - where the accretion and feedback actually occur - is still *terra incognita* because it has such low density and because UV-bright QSOs behind nearby galaxies are rare.

COS has recently opened more territory to explorations of the CGM by increasing the number of accessible QSOs about 20-fold. The COS-Halos survey (Tumlinson et al. 2011) has exploited this capability to systematically survey the halos of 50 L* galaxies at z ~ 0.2 with one QSO each. We have used the O VI λλ1032,1038 doublet to show that the CGM of star-forming galaxies contains as much oxygen as their dense ISM, with major consequences for galactic feedback and chemical evolution. We have also found that the halos of "red and dead" galaxies have a surprisingly large amount of cold, bound gas in their halos, indicating that quenching of star formation does not totally remove the halo gas reservoir. O VI and EUV lines such as Ne VIII and Mg X (redshifted into the LUV) are better for these studies than tracers at longer wavelengths (such as CIV) because they probe higher temperatures and are less likely to come from cool photoionized gas.

The LUV wavelength range was closed to COS-Halos, so it targeted galaxies at z ≳ 0.12 to place O VI λλ1032,1038 at > 1140 Å, the shortest wavelength available with good sensitivity given the HST primary mirror coating. The price of this constraint appears in Figure 2, which shows the parent sample of QSO/galaxy pairs from which the COS-Halos sample was drawn. The redshift $z > 0.11$ required to place the O VI doublet into the COS band means that galaxies *closer* than 500 Mpc are not observable in O VI, and the entire SDSS spectroscopic survey is *off limits* to this critical diagnostic! Owing to sensitivity and wavelength coverage, HST and FUSE can access only a tiny fraction of the CGM gas that could be observed if we had access to fainter QSOs and the LUV. This science also requires high spectral resolution (R > 50,000) to resolve the cold gas associated with low-metallicity CGM clouds and other ISM lines in different environments.

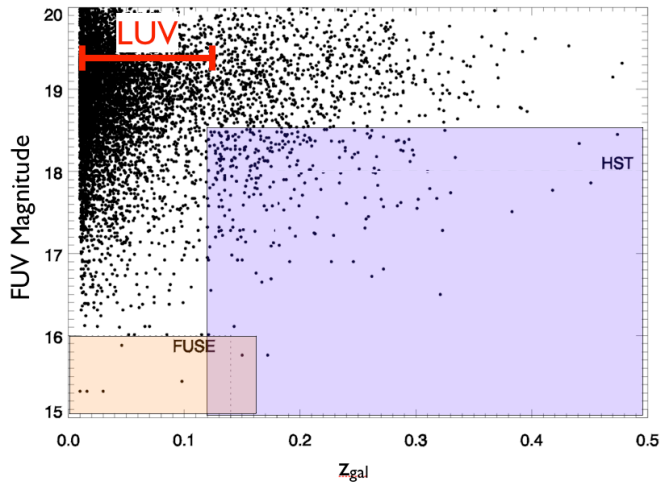
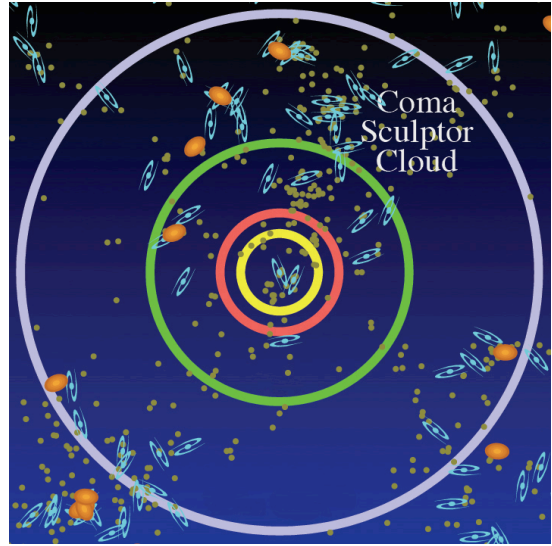

Figure 2: FUV accessibility to O VI, the key tracer of highly ionized CGM gas. Because of their limited sensitivity and wavelength coverage, HST and FUSE are still only able to probe a small fraction of the existing QSO galaxy pairs (black dots) that could be used to probe the CGM.

Figure 3: The outer circle is region of the local Universe where we can achieve > 10 QSOs behind every galaxy for highly resolved probes of the CGM with a telescope 100x HST sensitivity. However, without the LUV, a region **100000x** the volume of this sphere is **closed** to the key physical diagnostics of CGM gas.

Figure 3 illustrates another discovery space that could be opened with a future mission. A UV spectrograph with 100x the sensitivity of COS, which would be viable on an 8-m telescope, could observe *>10 QSOs behind every galaxy* out to 10 Mpc (the light purple) to weigh their CGM and relate its inflow and outflow to the stars and structure of fully resolved galaxies. Where COS-Halos is limited to studying mainstream L* galaxies, a future survey could effectively make a movie of the mass, metallicity, kinematics of the CGM in every phase of galaxy evolution from starbursts, to AGN, to passive evolution. A next-generation spectrograph could trigger another revolution in our understanding of the gas around galaxies and its role in galaxy formation. But without the LUV, this discovery space will be closed and a volume 100,000 times larger than the region of Figure 3 would still be inaccessible to the most important gas tracers.

### **Chemical Feedback in Galaxies**

Star-forming galaxies (SFGs) are characterized by a large reservoir of HI (≤ 90-95% of the baryonic matter, Kniazev et al. 2000), which can hide the bulk of the interstellar metals. Measuring the metal content in this gas phase is critical for a proper accounting of all the baryons and the LUV wavelength range is key to this census. Massive stars of type O and B can be used very effectively in the Local Volume to probe abundances in the ISM and gas flows in and out of the disk. With ~ 10-20 OB stars per kpc$^2$ in a star-forming galaxy like the LMC or larger, the distribution of the metals on local scales of the order of 100 pc can be mapped and correlated with the local SFR.

Access to the LUV is critical for these studies. First, it allows access to the the full Lyman series for accurate measurements of total gas content, and it covers the optimal oxygen transitions for measuring gas-phase abundances. However, oxygen measurements in the neutral ISM are severely compromised by saturation in the strong O I λ1302 line and the near total absence of other O lines strong enough to detect in the FUV. Access to the LUV gives access to many O transitions of varying strength that would allow a more accurate estimate of the O abundances in the neutral gas for direct comparison with HII regions.

Furthermore, the LUV added to 1200-3200 Å gives access to all the gas phases, including the cold molecular gas traced by $H_2$ (100 K), hot/coronal gas traced by OVI ($10^{5-6}$ K), warm gas ($10^4$ K) due to shocks/photoionization, and cold neutral interstellar medium (<1000 K). If the LUV is accessed at 100x greater sensitivity than COS at ~ 1300 Å, an OB star with the typical flux that is detected at S/N ~ 10 in ~ 50 HST orbits at the edge of the Local Group (1 Mpc), could actually be detected in **all** SFGs within a 10 Mpc Local Volume (as in Figure 3). **With IFU or ~100-fold multi-object capability, we would achieve a 1000- or 10,000-fold gain over HST that would allow us to map all the gas phases of the ISM at 100 pc scales in all star-forming galaxies within a 10 Mpc Volume.** This would allow us to understand how metals are released and recycled into galaxies of different type (normal L* galaxies vs LMC-like irregulars or dwarfs) and to have a comprehensive picture of chemical feedback in the Local Universe.

## Active Galactic Nuclei, Near and Far

Understanding how black holes accrete matter, grow through cosmic time, and influence their host galaxies is crucial for our understanding of galaxy evolution. The accretion disks powering most active galactic nuclei (AGN) emit most of their energy in the far and extreme ultraviolet energy range. Outflows from AGN, visible as blue-shifted ultraviolet and X-ray absorption lines from highly ionized species (Crenshaw et al. 2003), may be at the heart of feedback processes that regulate the growth of the host galaxy. For nearby (z < 0.15) AGN, the LUV band contains key spectral diagnostics (O VI and the Lyman lines) that let us measure the kinematics and abundance of outflowing gas from AGN. When combined with the doublets of C IV and N V, these enable the measurement of absolute abundances (Arav et al. 2007). The capability to cover the full 900—3200 Å band at COS sensitivities would enable such measurements for hundreds of AGN, permitting direct measurement of the absolute abundances of gas ejected by AGN into their host galaxies. These studies can be enabled using the same sample of hundreds of AGN chosen for studies of the CGM, described above.

At 0.2 < z < 2.0, lines of Ne VIII, Na IX, Mg X, and Si XII fall in the 900—3200 Å band. These ions have ionization potentials comparable to the X-ray absorbing gas detected in bright, local AGN. They have currently only been seen in the brightest intermediate-redshift AGN (Telfer et al. 1998; Muzahid et al. 2012). An LUV spectrograph with R~20,000 and a throughput of 5x COS would enable the detailed kinematical study of these species in hundreds of AGN at z > 0.2 more sensitively than any proposed X-ray telescope.

The peak of the spectral energy distribution of most AGN is in the extreme ultraviolet (Shang et al. 2011). While thermal emission from an optically thick accretion disk forms the peak of the spectrum, the shape in the extreme ultraviolet and how this connects to the soft X-ray is largely unknown due to absorption by neutral hydrogen and helium in the Milky Way. At intermediate redshifts, this spectral region becomes directly visible; observations in the LUV band open more of the AGN SED to direct observation. AGN accretion disks that peak at ~1200 Å (Telfer et al. 2003) are too cool for thermal radiation to continue to the soft X-ray band (e.g., Done et al. 2012). Comptonization of the disk spectrum by a warm, ionized coronal layer can produce the soft X-ray excess. Direct observation of this portion of the spectrum in intermediate redshift AGN (z~1) and correlation with the longer-wavelength thermal continuum to study time lags associated with the Comptonized reprocessing would enable us to assess the geometry of the accretion disk in hundreds of AGN. Opening the LUV also allows observations of QSO rest-frame EUV continua, as well as strong EUV emission lines (Ne VIII, OII-V) that have been seen in QSO composite spectra with COS (Shull, Stevans, & Danforth 2012).

### Lyman Continuum (LyC) at Low Redshift and the Reionization of the IGM

We know most of the universe is ionized, but how did it become so? The answer is shrouded by the extreme opacity of neutral H to the first ionizing sources that emerge in the epoch of reionization. Observations by WMAP and SDSS have bounded this epoch to between 0.4 - 1 Gyr after the Big Bang ($6 < z < 12$). A key goal of the JWST is to detect the sources responsible for reionization, which are thought to be large numbers of small star-forming galaxies. Their existence has been inferred from theoretical calculations, comparing the density of hydrogen to the density of ionizing photons emitted by the first stars. A fundamental parameter in these calculations is the fraction of ionizing photons that escape from star-forming galaxies. JWST cannot directly detect ionizing photons, due to the prevalence of Lyman Limit (LL) systems in the IGM with neutral hydrogen column densities in excess of $10^{17}$ cm$^{-2}$. Inoue & Iwata (2008) calculate that at $z = 5.8$ the probability is near zero for finding a line-of-sight with source transmission > 0.02 in the LyC and at $z = 3$ the probability is only 50% for line-of-sight transmission to be > 0.6.

McCandliss et al. (2008) argue that the sweet spot for directly detecting the escape of ionizing radiation from star-forming galaxies is at the lowest redshifts, below $z < 0.4$ where the Lyman edge lies between its rest value at 912 and 1276 Å. There the need to correct for attenuation by LL systems becomes vanishingly small. Moreover, the characteristic brightness of the star-forming population of galaxies is 4.5 magnitudes, nearly a factor of 100, higher at $z \sim 0.1$ than it is at $z = 3$. McCandliss et al. (2008) find, assuming an ionizing radiation escape fraction of 0.02, that the estimated characteristic brightness of the star-forming galaxy population in the Lyman continuum (LyC) is $10^{-17}$ erg cm$^{-2}$ s$^{-2}$ Å$^{-1}$ (~ 25 AB magnitude), at $z = 0.1$, while it is $10^{-19}$ erg cm$^{-2}$ s$^{-2}$ Å$^{-1}$ (~ 29.5 AB magnitude) at $z = 2$. Low-z observations of ionizing radiation escape from small star-forming galaxies, analogous to those responsible for reionization, can probe the escape faction parameter to the lowest levels, in a variety of spatially resolved regions, allowing insight into those environments that favor escape. Access to 912 Å instead of just 1150 Å opens up many more AGN targets at $z = 2-2.8$ that probe the later epoch of He II reionization driven by QSOs. Such observations will provide quantified constraint on the contributions of star-forming galaxies to the meta-galactic ionizing background across all epochs and confirm whether star-forming galaxies reionized the universe or not.